\documentclass{cmspaper}
\usepackage{graphicx}
\usepackage{rotate}
\usepackage{rotating}
\usepackage{relsize}
\usepackage{lineno}
\usepackage{slashed}
\usepackage{slashbox}
\usepackage{url}
\usepackage{amsmath}
\usepackage{longtable}
\newcommand{\lhc}{{LHC}}
\newcommand{\atlas}{{ATLAS}}
\newcommand{\cms}{{CMS}}
\newcommand{\cteq}{{\tt CTEQ6.6}}
\newcommand{\mstw}{{\tt MSTW2008}}
\newcommand{\gl}{\tilde{g}}
\newcommand{\sq}{\tilde{q}}
\newcommand{\sqb}{\tilde{q}^*}
\newcommand{\st}{\tilde{t}}
\newcommand{\stb}{\tilde{t}^*}
\newcommand{\sbot}{\tilde{b}}
\newcommand{\sbotb}{\tilde{b}^*}
\newcommand{\si}{\sigma}
\newcommand{\NLL}{\mathrm{NLL}}

\newcommand{\NLO}{\mathrm{NLO}}
\newcommand{\rhohat}{\hat{\rho}}
\newcommand{\shat}{\hat{s}}

\newcommand{\pdfctequp}{\ensuremath{\mathrm{PDF}_{\mathrm{CTEQ, up}}}}
\newcommand{\pdfcteqdown}{\ensuremath{\mathrm{PDF}_{\mathrm{CTEQ, down}}}}
\newcommand{\scactequp}{\ensuremath{\mathrm{SCA}_{\mathrm{CTEQ, up}}}}
\newcommand{\scacteqdown}{\ensuremath{\mathrm{SCA}_{\mathrm{CTEQ, down}}}}
\newcommand{\pdfmstwup}{\ensuremath{\mathrm{PDF}_{\mathrm{MSTW, up}}}}
\newcommand{\pdfmstwdown}{\ensuremath{\mathrm{PDF}_{\mathrm{MSTW, down}}}}
\newcommand{\scamstwup}{\ensuremath{\mathrm{SCA}_{\mathrm{MSTW, up}}}}
\newcommand{\scamstwdown}{\ensuremath{\mathrm{SCA}_{\mathrm{MSTW, down}}}}
\newcommand{\alphasup}{\ensuremath{\alpha_{\mathrm{S, up}}}}
\newcommand{\alphasdown}{\ensuremath{\alpha_{\mathrm{S, down}}}}

\newcommand{\alphas}{\ensuremath{\alpha_{\mathrm{s}}}}
\newcommand{\cteqnom}{\ensuremath{\mathrm{CTEQ}_{\mathrm{nom}}}}
\newcommand{\mstwnom}{\ensuremath{\mathrm{MSTW}_{\mathrm{nom}}}}
\newcommand{\ctequp}{\ensuremath{\mathrm{CTEQ}_{\mathrm{up}}}}
\newcommand{\cteqdown}{\ensuremath{\mathrm{CTEQ}_{\mathrm{down}}}}
\newcommand{\mstwup}{\ensuremath{\mathrm{MSTW}_{\mathrm{up}}}}
\newcommand{\mstwdown}{\ensuremath{\mathrm{MSTW}_{\mathrm{down}}}}

\def\nn{\nonumber}

\begin{document}

%
\begin{titlepage}

\flushright{CERN-PH-TH/2012-163}
\title{Supersymmetry production cross sections in $pp$ collisions at $\sqrt{s} = 7$ TeV}
 \begin{Authlist}
 Michael Kr\"amer
\Instfoot{rwth}{ Institut f\"ur Theoretische Teilchenphysik und Kosmologie, RWTH Aachen, Germany}
 Anna Kulesza
 \Instfoot{unims}{ Institut f\"ur Theoretische Physik, Westf\"alische Wilhelms-Universit\"at M\"unster, D-48149 M\"unster, Germany}
Robin van der Leeuw
\Instfoot{nikhef}{Nikhef National Institute for Subatomic Physics and University of Amsterdam, Amsterdam, Netherlands}
Michelangelo Mangano
 \Instfoot{cern}{European Organization for Nuclear Research, CERN, Switzerland}
Sanjay Padhi
 \Instfoot{ucsd}{University of California, San Diego, USA}
Tilman Plehn
 \Instfoot{heidelberg}{Institut f\"ur Theoretische Physik, Universit\"at Heidelberg, Germany}
Xavier Portell 
 \Instfoot{cern}{European Organization for Nuclear Research, CERN, Switzerland}
 \end{Authlist}

\begin{abstract}
This document emerged from work that started in January 2012 as a joint effort  by the ATLAS, CMS and LPCC 
supersymmetry (SUSY) working groups to compile state-of-the-art cross section predictions for SUSY particle 
production at the LHC. We present cross sections for various 
SUSY  processes in pp collisions at $\sqrt{s} =7$~TeV, including an estimate of the theoretical 
uncertainty due to scale variation and the parton distribution
functions. Further results for higher LHC centre-of-mass energies will be collected at https://twiki.cern.ch/twiki/bin/view/LHCPhysics/SUSYCrossSections. For squark and gluino production, which 
dominate the inclusive SUSY cross section, we employ calculations which include the resummation of soft gluon emission at  
next-to-leading logarithmic accuracy, matched to next-to-leading order (NLO) SUSY-QCD. In all other
cases we rely on NLO SUSY-QCD predictions. 
\end{abstract}
\end{titlepage}

\clearpage

\section{Introduction}
\label{sec:intro}
The search for supersymmetry (SUSY) is a central activity of the \lhc\ physics programme.
In the framework of the Minimal Supersymmetric extension of the Standard Model (MSSM) with R-parity conservation, SUSY particles are pair produced.
At the \lhc, the most copiously produced SUSY particles are expected to be the strongly interacting partners of quarks, the squarks ($\sq$), 
and the partners of gluons, the gluinos ($\gl$).  The squark and gluino pair-production processes to be considered are 
\begin{equation}
pp \to \sq\sq, \sq \sqb, \sq\gl, \gl\gl + X \,,
\label{eq:prod}
\end{equation}
together with the charge conjugated processes.  In Eq.~(\ref{eq:prod})
the chiralities of the squarks, $\sq=(\sq_L,\sq_R)$, are suppressed,
and we focus on the production of the partners of the $(u,d,c,s,b)$
quarks which we assume to be mass-degenerate. The production of the
SUSY partners of top quarks, the stops $(\st)$, and, when appropriate,
the partners of bottom quarks, the sbottoms $(\sbot)$, has to be
considered separately due to parton distribution function (PDF) effects and potentially large mixing
affecting the mass splittings. In this case, we explicitly specify the
different mass states in the pair-production processes,
\begin{equation}
pp \to \st_i \stb_i, \sbot_i \sbotb_i + X \qquad\qquad i=1,2\,,
\label{eq:stsb_prod}
\end{equation} 
where $i=1,2$ corresponds to the lighter and heavier states, respectively.  Furthermore, if weak-scale SUSY is realised in nature, the production of 
SUSY partners of the electroweak particles in the Standard Model, namely the gauginos and higgsinos, need also to be considered.

A comprehensive search program sensitive to the production of SUSY particles has been underway at the \lhc, since the beginning of the data-taking process.
These searches aim for a broad range of possible final states \cite{ATLAS_wiki, CMS_wiki}. 
Given the importance of SUSY searches at the \lhc, 
accurate knowledge of theoretical predictions for the cross sections is required.  
The interpretation, in terms of exclusion limits, of the results from the Tevatron, the 2010 \lhc\ 
data and the first set of the 2011 \lhc\ events, used next-to-leading order (NLO) predictions. However, over the past few years, 
substantial progress has been made in calculating higher-order corrections for squark and gluino production cross sections. 
In particular, the resummed results at the next-to-leading logarithmic (NLL) accuracy matched to NLO predictions have become available, allowing for a 
significant reduction of the theoretical uncertainties related to unknown higher perturbative orders.  
Therefore, it is desirable to implement this recent progress in the forthcoming set of analyses of the most recent LHC data sets.
In the case of electroweak or strong-weak production, i.e. production processes in which sleptons and gauginos are involved, 
the NLO predictions are used.

This document emerged from discussions among the SUSY working groups
of ATLAS, CMS and of the LHC physics center at CERN (LPCC)~\cite{lpcc}. The aim is to provide a reference for the collaborations on the evaluation
of the production cross sections and their associated uncertainties,
along the lines of similar recommendations developed by the {\sc
  PDF4LHC}~\cite{pdf4lhc} forum and by the Higgs cross section working
group~\cite{Dittmaier:2011ti,Dittmaier:2012vm}. 
For the purpose of illustration, this document will present some
explicit results for $\sqrt{s}=7$~TeV. The detailed cross section values for the
relevant processes and SUSY models considered by the
experiments, as well as the results for higher LHC centre-of-mass
energies, 
will be collected at the SUSY cross section working group 
twiki page~\cite{combined7TeV}.

The next section briefly describes the current state-of-the-art higher
order calculations, followed by the prescription used for the
treatment of theoretical uncertainties in section~\ref{sec:uncert}. In
sections~\ref{sec:xsectplots} and~\ref{sec:ewkinos}, the production
cross sections for coloured SUSY states and electroweak SUSY
productions are shown, respectively. A summary of the results and of the 
future prospects is given in section~\ref{sec:summary}.

\section{Higher order calculations -- NLO+NLL}
\label{sec:highorder}

The dependence of hadron collider observables on the renormalisation
and factorisation scales is an artifact of perturbation theory. 
This is typically reduced as higher-order perturbative contributions
are included. 
Assuming that there is no systematic shift of an
observable from order to order in perturbation theory, for example due
to the appearance of new production channels, the range of rates
covered by the scale
dependence at a given loop order should include
the true prediction of this rate. 
The scale dependence gives us therefore a lower limit on the theory
uncertainty of a QCD prediction, which becomes smaller as higher-order
SUSY-QCD corrections are included.
To estimate the scale uncertainty in this study we 
vary simultaneously factorisation and renormalisation scales, within a
range of 0.5 to 2 times the reference central scale $\mu$, where 
$\mu$ is the average of the two sparticle masses in the final state.
  In the
future, we shall study the effect of varying renormalisation and
factorisation scales independently. 

The corrections often increase the size of
the cross section with respect to the leading-order prediction
\cite{Kane:1982hw, Harrison:1982yi,Dawson:1983fw} if the
renormalisation and factorisation scales are chosen close to the
average mass of the produced SUSY particles. As a result, the SUSY-QCD
corrections have a substantial impact on the determination of mass
exclusion limits and would lead to a significant reduction of uncertainties
on SUSY mass or parameter values in the case of discovery, see e.g.\ \cite{Beenakker:2011dk}.
The
processes listed in Eqs.~(\ref{eq:prod}) and ~(\ref{eq:stsb_prod})
have been known for quite some time at NLO  in
SUSY-QCD~\cite{Beenakker:1994an,Beenakker:1995fp, Beenakker:1996ch,
  Beenakker:1997ut}. Note that SUSY-QCD corrections can be split into
two parts: first, the QCD corrections induced by gluon or quark
radiation and by gluon loops, which follow essentially the
same pattern as, for example, the top pair production, and second, the virtual 
diagrams which involve squark and gluino loops, which are independent of the real emission corrections. 
For relatively heavy
SUSY spectra the latter are numerically sub-leading terms which are,
however, 
challenging to compute. This is mainly due to a large number of Feynman diagrams 
 with many different mass scales contributing to the overall cross section.
For example for stop pair production, where neither light-flavour squarks nor gluinos
appear in the tree-level diagrams, these contributions can be easily
decoupled~\cite{Beenakker:1997ut}. The only part which requires some
attention is the appropriate treatment of the counter term and the
running of the strong coupling constant. This decoupling limit is
implemented in {\sc Prospino2}~\cite{prospino2}. For light-flavour squark and gluino
production this decoupling would only be consistent if applied to the
leading order as well as NLO contributions. This is usually not required, 
unless we choose specific simplified models.

Given the expected squark flavour structure in the MSSM, most
numerical implementations, including {\sc Prospino2}, make assumptions
about the squark mass spectrum. The left-handed and right-handed
squarks of the five light flavours are assumed to be mass
degenerate. Only the two stop masses are kept separate in the NLO
computations of light-flavour production
rates~\cite{Beenakker:1994an,Beenakker:1995fp, Beenakker:1996ch}. In
the {\sc Prospino2}~\cite{prospino2} implementation, this degeneracy is not assumed for
the leading-order results. However, the approximate NLO rates are
computed from the exact leading order cross sections times the mass
degenerate $K$ factors. For the pair production of third-generation
squarks the four light squark flavours are mass degenerate, while the
third generation masses are kept
separate~\cite{Beenakker:1997ut}. This approximation can for example
be tested using {\sc MadGolem}~\cite{Binoth:2011xi}, an automized NLO
tool linked to {\sc MadGraph4}~\cite{Alwall:2007st}. It is also important 
to point out here that in {\sc Prospino2} the pair production of
third-generation squarks is available as individual
processes. However, sbottom pairs are included in the implicit sum of
light-flavour squarks because there is no perfect separation of bottom
and light-flavour decay jets.

When summing the squark and gluino production rates including
next-to-leading order corrections it is crucial to avoid double
counting of processes. For example, squark pair production includes
$\mathcal{O}(\alphas^3)$ processes of the kind $qg \to \sq \sq^*q$. 
The same final state can be produced in $\sq \gl$ production when
the on-shell gluino decays into an anti-squark and a quark. The {\sc
  Prospino} scheme for the separation and subtraction of on-shell
divergences from the $\sq \sq^*$ process uniquely ensures a consistent
and point-by-point separation over the entire phase space. For a
finite particle mass it has recently been adopted by {\sc MC\@@NLO}~\cite{Weydert:2009vr}
for top quark processes. It is automized as part of {\sc
  MadGolem}~\cite{Binoth:2011xi}.

A significant part of the NLO QCD corrections can be attributed to the
threshold region, where the partonic centre-of-mass energy is close to
the kinematic production threshold. In this case the NLO
corrections are typically large, with the most significant
contributions coming from soft-gluon emission off the coloured
particles in the initial 
and final state. The contributions due to soft gluon emission can be
consistently taken into 
account to all orders by means of threshold resummation. In this
paper, we discuss results where resummation has been performed at 
next-to-leading logarithmic (NLL) accuracy~\cite{Kulesza:2008jb,Kulesza:2009kq,
Beenakker:2009ha,Beenakker:2010nq,Beenakker:2011fu}. 

The step from NLO to NLO+NLL is achieved by calculating the NLL-resummed partonic cross section 
$\tilde \sigma^{\rm (NLL)}$ and then matching it to the NLO prediction, in order to retain the available information 
on other than soft-gluon contributions. The matching procedure takes the following form
\begin{eqnarray}
\label{eq:matching}
\si^{\rm (NLO+NLL)}_{p p \to kl}\bigl(\rho, \{m^2\},\mu^2\bigr) 
  &=& \si^{\rm (NLO)}_{p p \to kl}\bigl(\rho, \{m^2\},\mu^2\bigr)\nn
          \\[1mm]
   &&  \hspace*{-30mm}+\, \frac{1}{2 \pi i} \sum_{i,j=q,\bar{q},g}\, \int_\mathrm{CT}\,dN\,\rho^{-N}\,
       \tilde f_{i/p}(N+1,\mu^2)\,\tilde f_{j/p}(N+1,\mu^2) \nn\\[0mm]
   && \hspace*{-20mm} \times\,
       \left[\tilde\si^{\rm(NLL)}_{ij\to kl}\bigl(N,\{m^2\},\mu^2\bigr)
             \,-\, \tilde\si^{\rm(NLL)}_{ij\to kl}\bigl(N,\{m^2\},\mu^2\bigr)
       {\left.\right|}_{\scriptscriptstyle({\NLO})}\, \right]\,,
\end{eqnarray}
where the last term in the square brackets denotes the NLL resummed
expression expanded to NLO. The symbol $\{m^2\}$ stands for all masses entering the 
calculations and  $\mu$ is the common factorisation and
renormalisation scale. The resummation is performed in the Mellin moment $N$ space, 
with all Mellin-transformed quantities indicated by tilde. In particular, the Mellin moments of the 
partonic cross sections are defined as 
\begin{equation}
\label{eq:Mellin}
  \tilde\si_{i j \to kl}\bigl(N, \{m^2\}, \mu^2 \bigr) 
 \equiv \int_0^1 d\rhohat\;\rhohat^{N-1}\;
           \si_{ i j\to kl}\bigl(\rhohat,\{ m^2\}, \mu^2 \bigr) \,.
\end{equation}
The variable $\rhohat \equiv (m_k + m_l)^2/\shat $ measures the closeness to the partonic production threshold 
and is related to the corresponding hadronic variable $\rho = \rhohat x_i x_j$ in Eq.~(\ref{eq:matching}), where $x_i\ (x_j)$ 
are the usual longitudinal momentum fraction of the incoming proton carried out by the parton $i (j)$. 
The necessary inverse Mellin transform in Eq.~(\ref{eq:matching}) is performed  along the contour ${\rm CT}$ according to the 
so-called ``minimal prescription''~\cite{Catani:1996yz}.
The NLL resummed cross section in
Eq.~(\ref{eq:matching}) reads
\begin{eqnarray}
  \label{eq:NLLres}
  \tilde{\sigma}^{\rm (NLL)} _{ij\rightarrow kl}\bigl(N,\{m^2\},\mu^2\bigr) 
& = &\sum_{I}\,
      \tilde\sigma^{(0)}_{ij\rightarrow
        kl,I}\bigl(N,\{m^2\},\mu^2\bigr)\, \nn  \\[1mm]
   & \times\,& \Delta^{\rm (NLL)}_i (N+1,Q^2,\mu^2)\,\Delta^{\rm (NLL)}_j (N+1,Q^2,\mu^2)\,
     \Delta^{\rm (s, \NLL)}_{ij\rightarrow
       kl,I}\bigl(N+1,Q^2,\mu^2\bigr)\,,
\end{eqnarray}
where the hard scale $Q^2$ is taken as $Q^2 = (m_k + m_l)^2$ and $\tilde{\sigma}^{(0)}_{ij \rightarrow kl, I}$ are the
colour-decomposed leading-order cross sections in Mellin-moment space,
with $I$ labelling the possible colour
structures.  The functions $\Delta^{\rm (NLL)}_{i}$ and $\Delta^{\rm (NLL)}_{j}$ sum
the effects of the (soft-)collinear radiation from the incoming
partons. They are process-independent and do not depend on the colour
structures.  These functions contain both the leading logarithmic as well
as part of the sub-leading logarithmic behaviour. The expressions for
$\Delta^{\rm (NLL)}_{i}$ and $\Delta^{\rm (NLL)}_{j}$ can be found in the
literature~\cite{Kulesza:2009kq}. In order to perform resummation at NLL accuracy, one also has to take 
into account soft-gluon contributions involving emissions from the final state, depending on the colour structures in which the
final state SUSY particle pairs can be produced. They are summarised by the factor
\begin{equation}
  \Delta_{I}^{\rm (s, \NLL)}\bigl(N,Q^2,\mu^2\bigr) 
  \;=\; \exp\left[\int_{\mu}^{Q/N}\frac{dq}{q}\,\frac{\alphas(q)}{\pi}
                 \,D_{I} \,\right]\,.
\label{eq:2}
\end{equation}
The one-loop coefficients $D_{I}$ follow from the threshold limit of
the one-loop soft anomalous-dimension
matrix and can be found in~\cite{Kulesza:2009kq,Beenakker:2009ha}. 

The analytic results for the NLL part of the cross sections have been implemented into a numerical code. 
The results of this code, added to the NLO results obtained from {\sc Prospino2}, correspond to the matched NLO+NLL cross sections. 
Their central values, the scale uncertainty and the 68\% C.L. pdf and $\alphas$ uncertainties obtained using \cteq~\cite{cteq66} and \mstw~\cite{mstw08}
PDFs have been tabulated for the squark and gluino production processes of interest in the range of input masses appropriate for the current 
experimental analysis.\footnote{\protect Note that we use NLO pdfs with the NLO+NLL matched cross section calculation. The reduction
of the factorisation scale dependence observed in the NLO+NLL predictions is a result of a better
compensation between the scale dependence of the NLO evolution of the pdf and the short
distance cross section, and does not depend on whether pdfs are fitted using NLO or NLO+NLL
theory, see for example Ref.~\cite{Sterman:2000pu}. In general, one can apply NLL
threshold resummation with
NLO pdfs to processes like heavy SUSY particle production for which the summation of logarithms
is more important than for the input data to the NLO fits. However, it would be interesting to systematically study the difference  
between NLO and NLO+NLL input to global pdf determinations for SUSY
particle production at the LHC.}
The set of tabulated values has enough granularity in order to minimise the interpolation uncertainty when an intermediate 
value is used as input in the analysis. Together with a corresponding interpolation code, the tabulated values constitute the {\sc NLL-fast} 
numerical package~\cite{Beenakker:2011fu,nllfast}.

Note that for specific processes, results beyond NLL accuracy are
already available. The production of squark-antisquark pairs has been 
calculated at next-to-next-to-leading-logarithmic (NNLL) level 
\cite{Beenakker:2011sf}. Also, a general
formalism has been developed in the framework of effective field
theories which allows for the resummation of soft and Coulomb gluons
in the production of coloured
sparticles~\cite{Beneke:2009rj,Beneke:2010da} and subsequently applied to squark and gluino production~\cite{Beneke:2010da, Falgari:2012hx}. 
In addition, the dominant next-to-next-to-leading order (NNLO) corrections, including
those coming from the resummed cross section at
next-to-next-to-leading-logarithmic (NNLL) level, have been calculated
for squark-antisquark pair-production
\cite{Langenfeld:2009eg,Langenfeld:2010vu}. 
The production of gluino
bound states as well as bound-state effects in gluino-pair and
squark-gluino production
has also been studied \cite{Hagiwara:2009hq,Kauth:2009ud,Kauth:2011vg,Kauth:2011bz}. 
Furthermore, electroweak corrections to the ${\cal O} (\alphas^2)$ tree-level
processes~\cite{Hollik:2007wf, Hollik:2008yi, Hollik:2008vm, Beccaria:2008mi,
    Mirabella:2009ap, Germer:2010vn, Germer:2011an} and the
electroweak Born production channels of ${\cal O} (\alpha\alphas)$
and ${\cal O} (\alpha^2)$~\cite{Alan:2007rp,Bornhauser:2007bf} are in
general significant for the pair production of SU(2)-doublet squarks
$\tilde{q}_L$ and at large invariant masses, but they are moderate for
inclusive cross sections and will not be included in the results
presented here. 

Resummation has also been studied for the production of electroweak
SUSY partcles, see Refs.~\cite{Bozzi:2006fw, Bozzi:2007qr, Li:2007ih,
  Bozzi:2007tea, Debove:2009ia, Debove:2010kf, Debove:2011xj, Broggio:2011bd}. However, since the electroweak
processes do not significantly contribute to the inclusive SUSY cross
section in general, we use the NLO-SUSY QCD calculation
\cite{Beenakker:1999xh, Berger:2000iu, Spira:2002rd} for their evaluation.

\section{Treatment of cross sections and their associated uncertainties}
\label{sec:uncert}

Here, the determination of the cross section central
values and associated systematic uncertainties is discussed in detailed,
in view of the recent progress discussed in the previous section.

In case of pair production processes of strongly interacting particles, for which NLL
calculations exist, the cross sections are taken at the
next-to-leading order in the strong coupling constant, including the
resummation of soft gluon emission at the NLL level of accuracy, performed using
the {\sc{NLL-fast}} code.

These cross sections are currently available for masses spanning
$200$~GeV to $2$~TeV for squarks and gluinos
and $100$~GeV to $1$~TeV for direct stop or sbottom pair
production. In the case of associated squark-gluino production and
gluino-pair production processes, the calculations are extended up to
squark masses of 3.5~TeV and 4.5~TeV, respectively. Following the
convention used in {\sc Prospino2}, in the case of
squarks, which can be more or less degenerate depending on a specific SUSY
scenario, the input mass used is the result of
averaging only the first and second generation squark masses. 
Further details on different scenarios considered to interpret the variety of
experimental searches developed by the \atlas\ and \cms\ collaborations
are described in Section~\ref{sec:special}.

In the case of other types of production processes, such as
electroweak production and strong-weak production, or in few cases
where the sparticle masses fall outside the range previously
described, the NLO approach is considered using {\sc Prospino2}.

A special treatment is performed in scenarios in which either the
squarks or the gluinos are set at a very large scale such that their
production is not possible at the LHC.  Defining this large scale is
arbitrary and in some cases it may have a non-negligible impact in the
production of the SUSY particles residing at the TeV scale
(e.g. squarks at high scales can still contribute to the gluino pair
production process via a $t$-channel exchange). Thus, instead of
setting an explicit arbitrary value for the mass of the decoupled
particle, the NLO+NLL calculation implemented in {\sc{NLL-fast}}
assumes that this production does not interfere in any possible way
with the production processes of the rest of the particles. Thus, the
particle is completely decoupled from the rest of the phenomenology at
the TeV scale.

The uncertainties due to the choice of the renormalisation and
factorisation scales as well as the PDF are obtained using the
{\sc{NLL-fast}} code or computed with {\sc Prospino2}.
In order to combine all these predictions and obtain an overall
uncertainty, the {\sc PDF4LHC} recommendations are followed as close
as possible, subject to the current availability of the different
calculations. Thus, an envelope of cross section predictions is
defined using the 68\% C.L. ranges of the \cteq~\cite{cteq66}
(including the $\alphas$ uncertainty) and \mstw~\cite{mstw08} PDF
sets, together with the variations of the scales. The nominal cross
section is obtained using the midpoint of the envelope and the
uncertainty assigned is half the full width of the
envelope. Mathematically, if \pdfctequp\ (\pdfcteqdown) and
\scactequp\ (\scacteqdown) are the upwards (downwards) one sigma
variations of the \cteq\ PDF set, respectively,
\pdfmstwup\ (\pdfmstwdown) and \scamstwup\ (\scamstwdown) are the
corresponding variations for the \mstw\ PDF set and, finally,
\alphasup\ (\alphasdown) is the corresponding up (down) one sigma
uncertainty of the \alphas\ coupling constant, the following
quantities can be calculated:
\begin{subequations}
\begin{align}
\label{eq:variations}
&\ctequp = \sqrt{\pdfctequp^2 + \scactequp^2+\alphasup^2} \,,\\
&\cteqdown = \sqrt{\pdfcteqdown^2 + \scacteqdown^2+\alphasdown^2}\,, \\
&\mstwup = \sqrt{\pdfmstwup^2 + \scamstwup^2} \,,\\
&\mstwdown = \sqrt{\pdfmstwdown^2 + \scamstwdown^2}\,.
\end{align}
\end{subequations}
\noindent The corresponding upper and lower values of the envelope created by this set of numbers and the nominal predictions (\cteqnom\ and \mstwnom) is obtained by:
\begin{subequations}
\begin{align}
\label{eq:UandL}
&\mathrm{U} = \mathrm{max} (\cteqnom + \ctequp, \mstwnom + \mstwup )\,, \\
&\mathrm{L} = \mathrm{min} (\cteqnom - \cteqdown, \mstwnom - \mstwdown )\,,
\end{align}
\end{subequations}
\noindent and the final corresponding cross section ($\sigma$) and its symmetric uncertainty ($\Delta\sigma$) is taken to be: 
\begin{subequations}
\begin{align}
\label{eq:sigmas}
\sigma &= (\mathrm{U}+\mathrm{L})/2 \,,\\
\Delta\sigma &= (\mathrm{U}-\mathrm{L})/2\,.
\end{align}
\end{subequations}
Full compliance with the PDF4LHC recommendations, with the inclusion
of other PDF sets such as {\sc NNPDF}~\cite{nnpdf}, will be
implemented in the near future. We notice that, as discussed in
section~\ref{sec:xsectplots}, the additional contribution to the
systematics coming from \alphas\ uncertainties is negligible.

\subsection{Special cases}
\label{sec:special}
Some models require special treatment in order to make sure that the
NLO cross sections are correctly computed.  This is important because
of the way higher order calculations handle different squark flavours,
which could lead to double-counting of diagrams. Given the difficulty
to provide a comprehensive summary of all situations that are
being considered in the interpretation of the LHC data, we only
discuss here few relevant cases, to exemplify the approach followed.

\subsubsection*{Simplified Models}

A variety of simplified models~\cite{Alves:2011wf} are considered by the experiments. In some cases, the gluino, 
sbottom, and stop squarks are decoupled from the rest of the
supersymmetric spectrum.\footnote{While this scenario is possible if we only
consider the TeV scale, it bears some challenges at higher energy
scales. Any kind of renormalisation scale evolution will generate
squark masses at the scale of the gluino mass, but not vice versa. 
Thus,  any discovery of light squarks associated with heavy
gluinos would point to a non-standard underlying
model~\cite{Jaeckel:2011wp}. In spite of all theory prejudice it is
clearly adequate that these regions be experimentally explored.}
In this specific simplified model, only squark-antisquark production
is allowed and this process is flavour-blind, if the masses are
considered degenerate. Since the NLO+NLL calculations consider the
sbottom as degenerate in mass with the squarks of the first and second
generations, the overall cross section has to be rescaled down by a
factor of $4/5=0.8$.

In other cases where the gluino is not decoupled, 
squark-gluino and squark-squark productions are feasible and they are not corrected further. 
The only effect could come from a $b$-quark in the initial state,
which is suppressed~\cite{Beenakker:2010nq}.
Other types of simplified models decouple not only the squarks from the third generation particles, but also all
the right-handed squarks.  These scenarios primarily focus on final state decays via charginos or neutralinos. 
The squark mass is calculated by averaging the non-decoupled squark masses and the final cross section is rescaled down by 
a factor of $(4/5)\cdot (1/2)=0.4$.

\subsubsection*{Treatment of 3rd generation squarks}
Direct stop and sbottom production must be treated
differently from the rest of squark families because, for instance,
the $t$-channel gluino-exchange diagrams are suppressed. 
In beyond the leading-order computations processes 
involving sbottoms can be degenerate with the rest of the other squark flavours, this
could lead to a potential double counting of sub-processes. 
In order to avoid this, in scenarios
in which the production of different squark flavours are present, the squark pair production cross
section is rescaled down to subtract the sbottom contribution and the
corresponding process is computed separately.

At leading order the production cross sections for the third-generation squarks 
depend only on their masses, and the results for sbottom and stop of
the same mass are therefore equal. At NLO in SUSY-QCD, additional SUSY
parameters like squark and gluino masses or the stop/sbottom mixing
angle enter. Their numerical impact, however, is very
small~\cite{Beenakker:1997ut, Beenakker:2010nq}. 
A further
difference between stop and sbottom pair production arises from the 
$b\bar{b} \to \tilde{b}\tilde{b}^*$ channel, where the initial-state
bottom quarks 
do allow a $t$-channel gluino-exchange graph that gives rise to extra
contributions. However, as has been demonstrated in Ref.~\cite{Beenakker:2010nq} their
numerical impact on the hadronic cross sections is negligible. Thus,
for all 
practical purposes, the LO and higher-order cross-section predictions 
obtained for stop-pair production apply also to sbottom-pair
production if the 
input parameters, i.e.\ masses and mixing angles, are modified accordingly.

\newcommand{\myxsectcaption}[2] {NLO+NLL {#1} production cross section
  with {#2} decoupled  as a function of mass. The different styled black (red) lines
  correspond to the cross section and scale uncertainties predicted
  using the \cteq\ (\mstw) PDF set. The yellow (dashed black) band
  corresponds to the total \cteq\ (\mstw) uncertainty, as described in
  the text. The green lines show the final cross section and its total
  uncertainty.}

\section{Gluino and squark production at the LHC}
\label{sec:xsectplots}

The production cross sections and associated uncertainties
resulting from the procedure described in the previous section are
discussed here for different processes of interest. 
We discuss three distinct cases:
gluino pair production with squarks decoupled, squark-antisquark pair
production with gluino decoupled and stop/sbottom pair production. The
results shown here are mainly illustrative: tables with cross sections and
systematics obtained in other scenarios which take into account the full complexity of
the spectrum, as a function of the parameters of various SUSY models,
are collected at the SUSY cross section working group twiki page~\cite{combined7TeV}.

\subsection{Gluino pair production}

\begin{figure}[htbp]
  \begin{center}
       	\includegraphics[width=12cm]{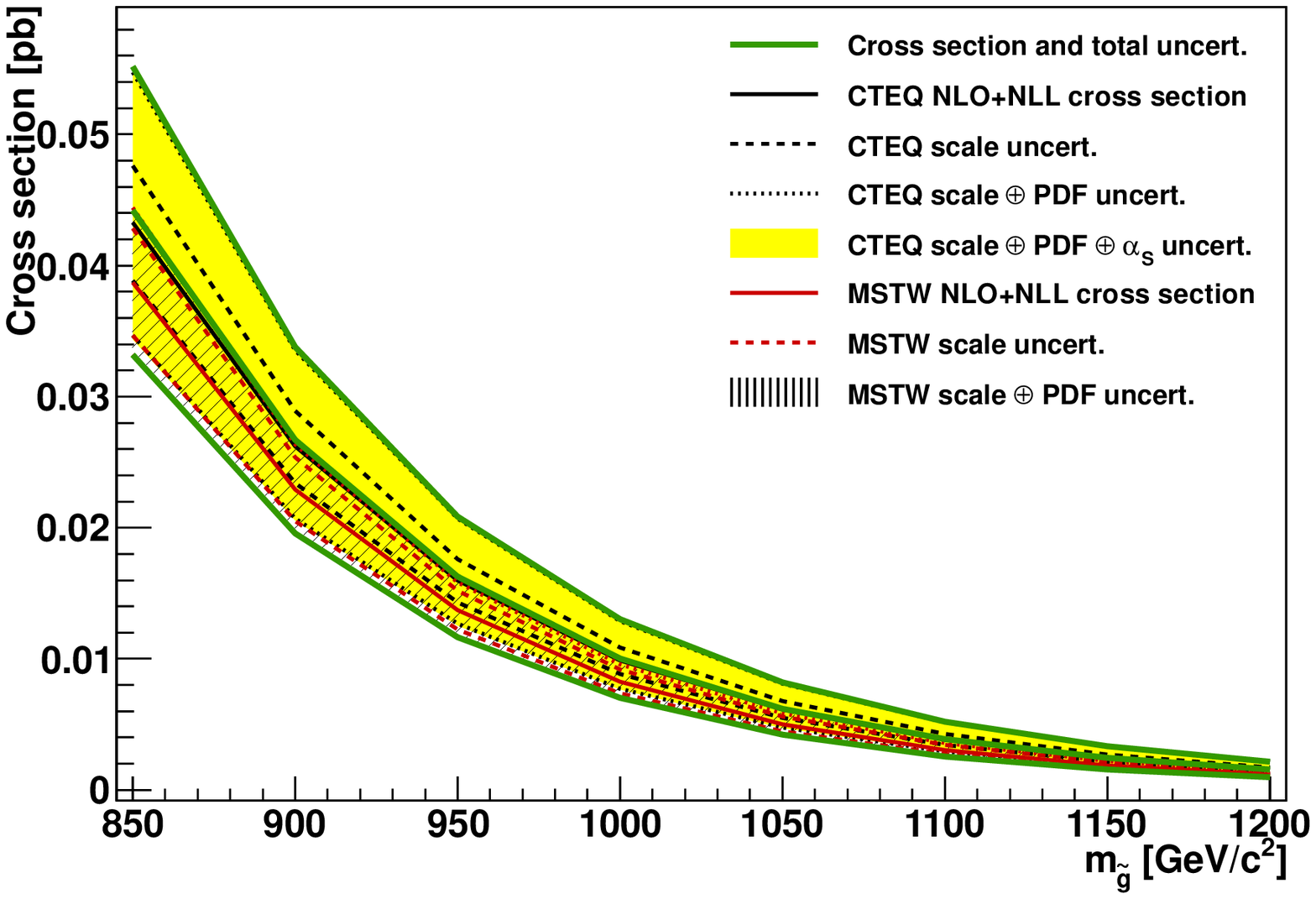}
  \end{center}
  \caption{\myxsectcaption{gluino pair}{squarks}}
  \label{fig:xsect_gluino}
\end{figure}

The gluino pair production cross section in a model where the squarks
are decoupled is shown in Figure~\ref{fig:xsect_gluino}, for a gluino
mass range spanning the current sensitivity of the \atlas\ and
\cms\ experiments. In the figure, the black (red) line corresponds to
the NLO+NLL nominal cross section and renormalisation and
factorisation scale uncertainties obtained using the \cteq\ (\mstw)
PDF set. The solid yellow (dashed black) band corresponds to the total
uncertainty of the cross section using \cteq\ (\mstw), as derived from
Eq.~\ref{eq:variations}. Finally, the green solid lines delimit the
envelope and the central value. They correspond to the central nominal
value along with the total uncertainties.

As it can be seen, the cross sections predicted using the \mstw\ PDF
set are lower than the ones predicted by \cteq.~For this mass range
the uncertainty using \mstw~is approximately half that of the
\cteq~set.  Thus, the impact of the \mstw\ prediction in the current
prescription is almost negligible. The small difference between the green and black dotted
lines also shows that the impact of the
\alphas\ uncertainty in the \cteq\ case is negligible. This is
consistent with the intuition that at the large
values of $x$ that are relevant for SUSY particle production, the
intrinsic PDF uncertainty dominates over the small evolution
differences caused at these high scales by the \alphas\ uncertainty.
This justifies as a good approximation not including the \alphas\ uncertainty
in the \mstw\ case.

\subsection{Squark-antisquark production}

\begin{figure}[htbp]
  \begin{center}
       	\includegraphics[width=12cm]{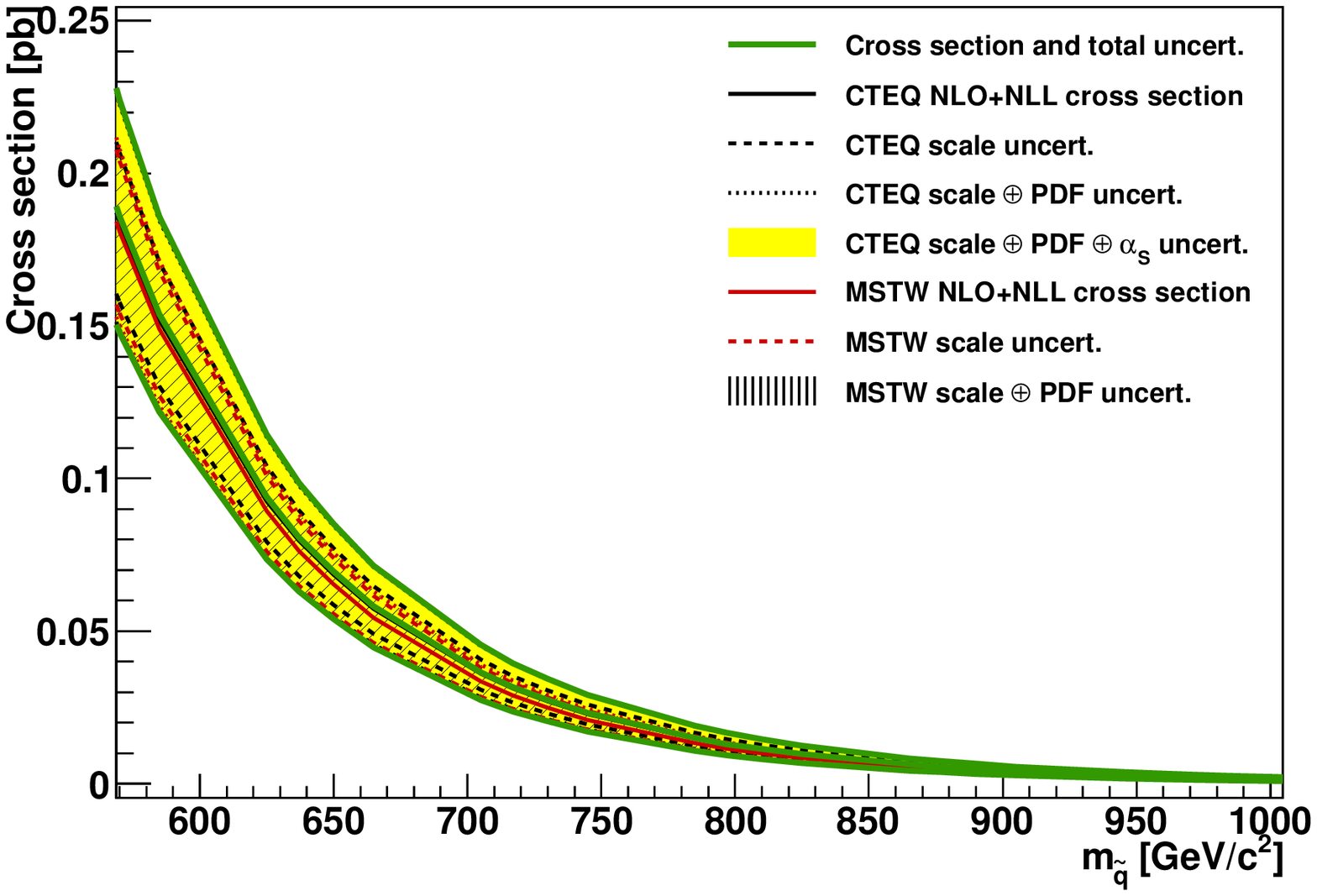}
  \end{center}
  \caption{\myxsectcaption{squark-antisquark}{gluinos}}
  \label{fig:xsect_squark}
\end{figure}

In order to show the evolution of the squark-antisquark production
cross section as a function of the squark mass, a scenario has been
chosen in which the gluino is decoupled. The results are shown in
Figure~\ref{fig:xsect_squark}, using the same graphical convention
described in the gluino pair production case.  The central values
predicted using the \cteq\ and \mstw\ PDF sets are relatively close to
each other. Again, the \mstw\ uncertainty is the smallest of the two,
although relatively larger than in case of gluino pair production.

\subsection{Direct stop and sbottom productions}

\begin{figure}[htbp]
  \begin{center}
       	\includegraphics[width=12cm]{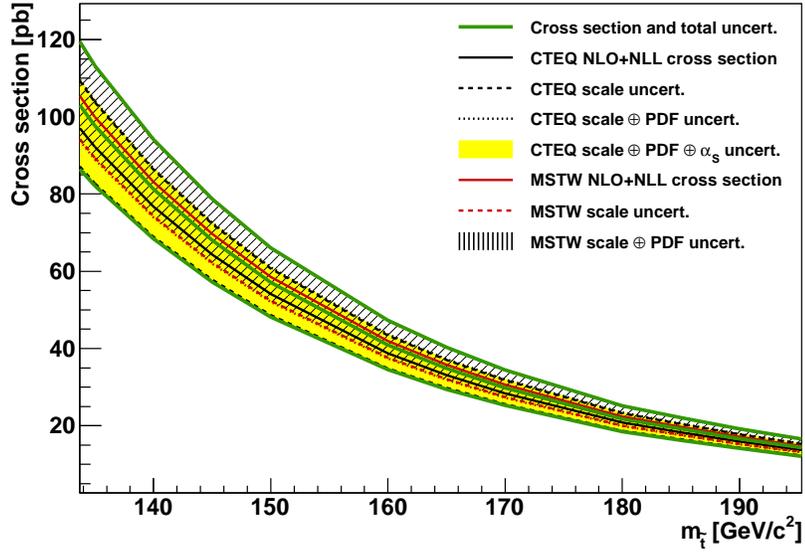}
       	\includegraphics[width=12cm]{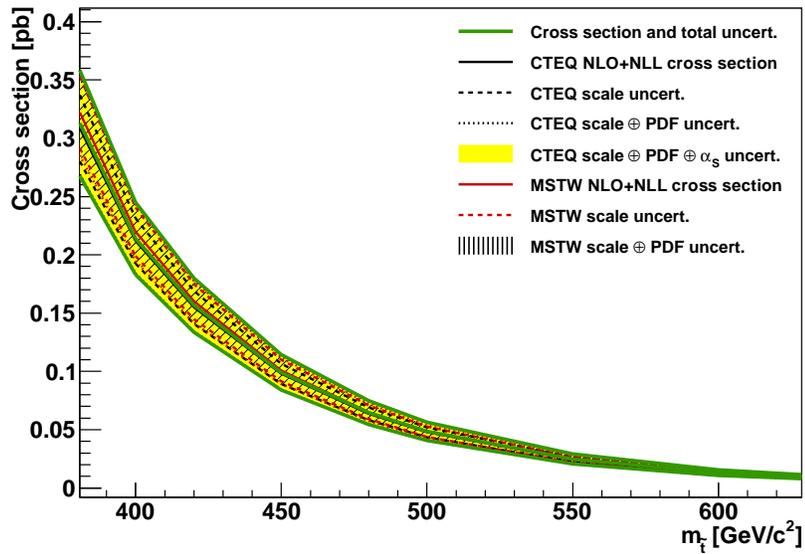}
  \end{center}
  \caption{NLO+NLL stop-antistop production cross section as a function of mass. The different 
   styled black (red) lines correspond to the cross section and scale uncertainties predicted
  using the \cteq\ (\mstw) PDF set. The yellow (dashed black) band
  corresponds to the total \cteq\ (\mstw) uncertainty, as described in
  the text. The green lines show the final cross section and its total
  uncertainty. The figure on the top (bottom) shows a low (high) stop mass range.}
  \label{fig:xsect_stop}
\end{figure}

The production cross section as a function of the stop mass for a
model in which only the lightest stop is reachable by the LHC is shown
in Figure~\ref{fig:xsect_stop}.  It should be noted that these cross
sections are the same to those of a model in which only the lightest
sbottom is accessible, assuming the rest of the coloured SUSY spectrum
decoupled.

We discuss two different cases depending on the stop mass.  For low
stop masses, the cross sections predictions using the \mstw\ PDF set 
are larger than the ones using \cteq, but the relative uncertainties
are similar. Thus, the
total cross section and the overall uncertainty are larger than those 
predicted by
\cteq\ alone. For larger stop
mass ranges, the predictions using any of the two PDF sets under
consideration are quite similar for the central value and its
associated uncertainties. The final prediction is mostly determined by the
\cteq\ PDF set.

\section{SUSY electroweak production at the LHC}
\label{sec:ewkinos}
\begin{figure}[htbp]
  \begin{center}
        \includegraphics[width=12cm]{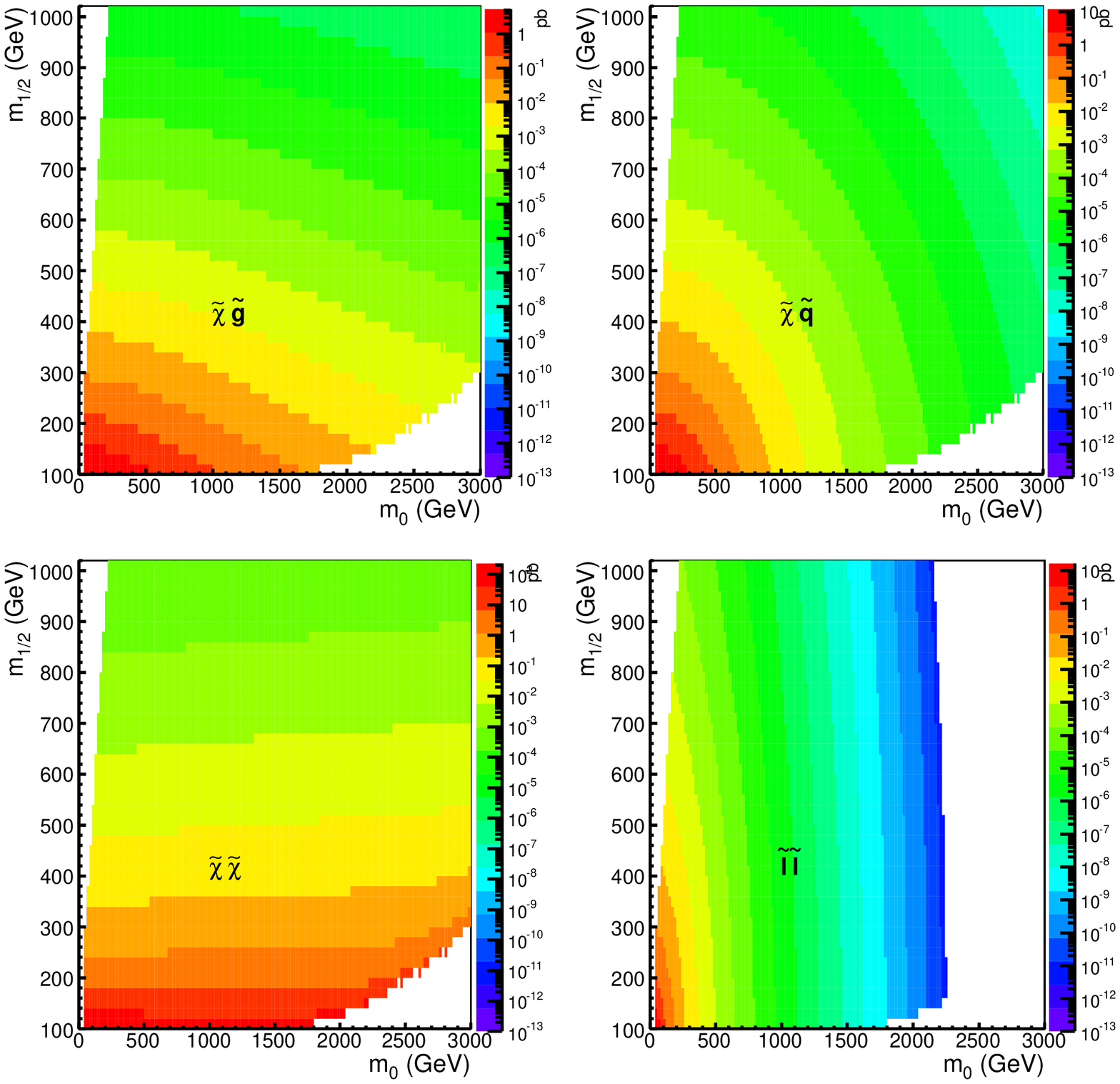}
  \end{center}
  \caption{NLO cross sections for SUSY electroweak productions in the
    $m_{0} - m_{1/2}$ plane using \cteq~PDF set.  The CMSSM framework is used, with
    $\tan\beta = 10, A_{0} = 0$~GeV and $ \mu > 0$. }
  \label{fig:xsect_ewknom}
\end{figure}

\begin{figure}[htbp]
  \begin{center}
        \includegraphics[width=12cm]{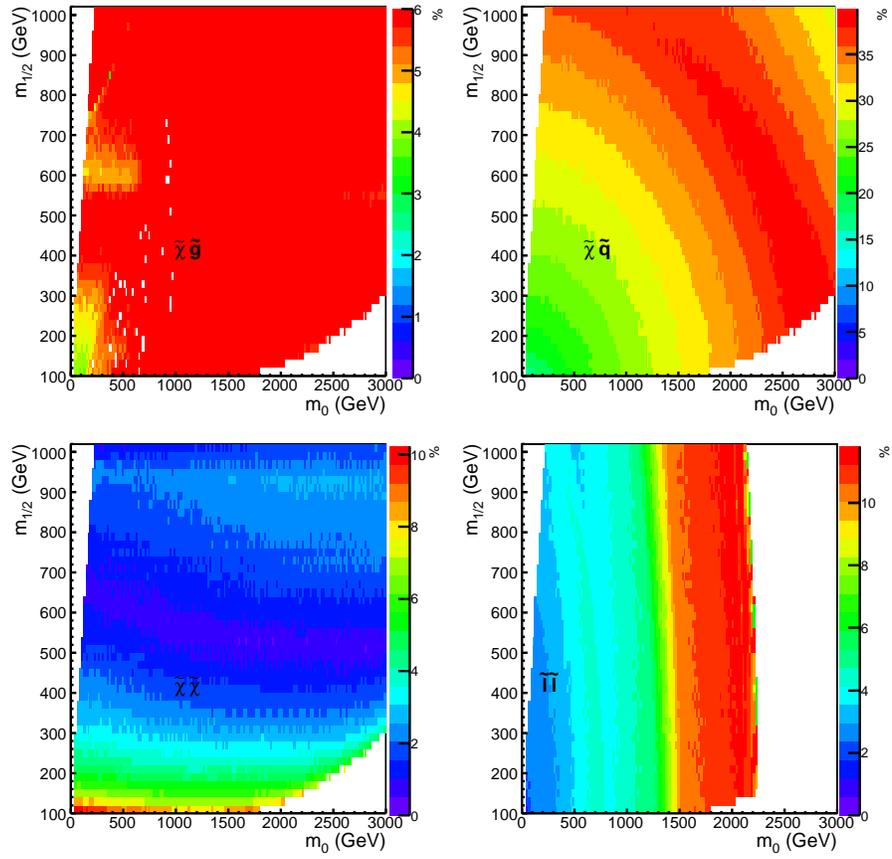}
  \end{center}
  \caption{Uncertainties in percentage due to the scale choice for
    SUSY electroweak productions in the $m_{0} - m_{1/2}$ plane using \cteq~PDF set. The
    CMSSM framework is used, with $\tan\beta = 10, A_{0} = 0 $~GeV and
    $ \mu > 0$. }
  \label{fig:xsect_ewkscale}
\end{figure}

\begin{figure}[htbp]
  \begin{center}
        \includegraphics[width=12cm]{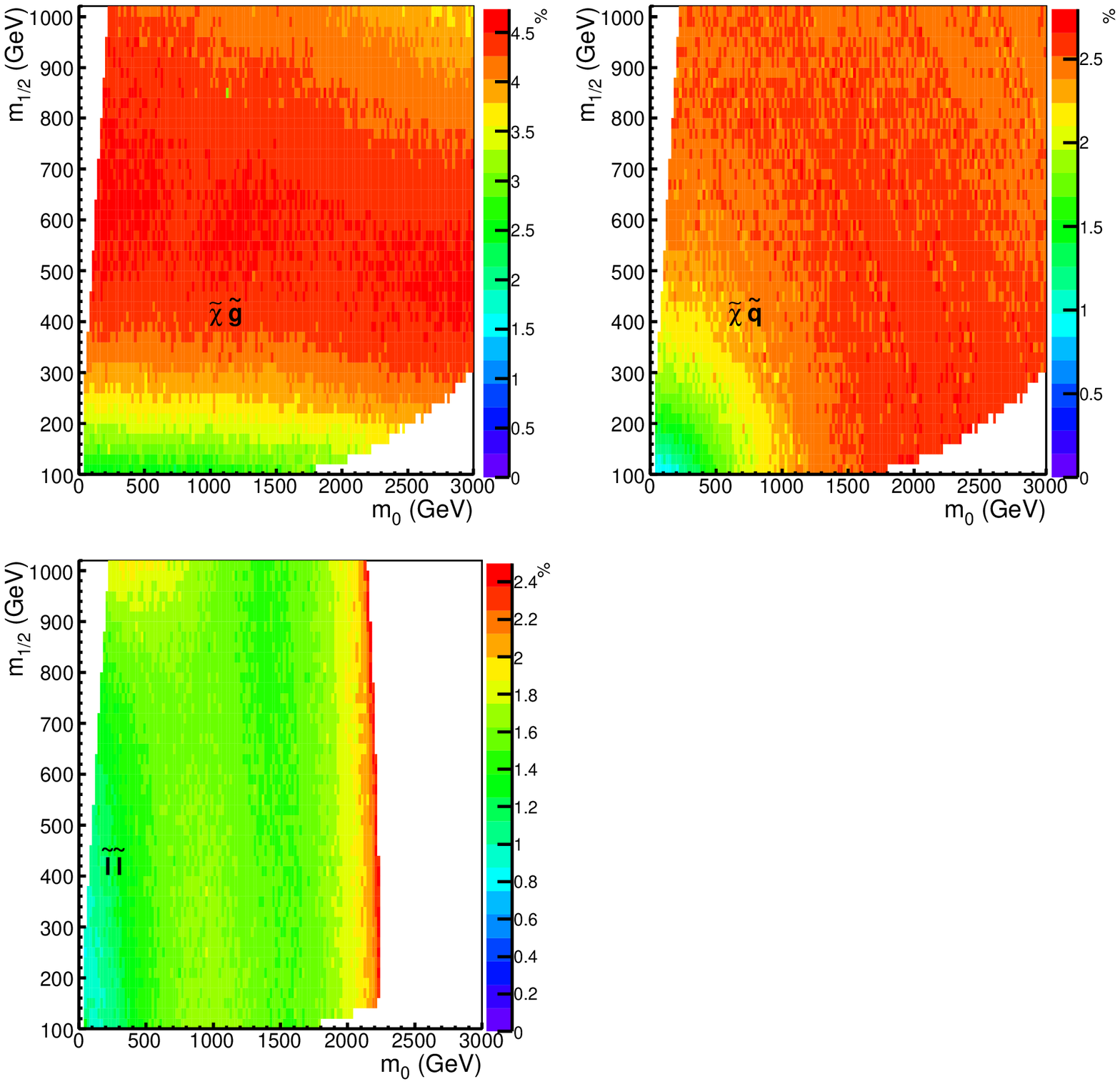}
  \end{center}
  \caption{Uncertainties in percentage due to \cteq~PDF error prescription for SUSY
    electroweak productions in the $m_{0} - m_{1/2}$ plane. The CMSSM
    framework is used with $\tan\beta = 10, A_{0} = 0$~GeV and $ \mu >
    0$. 
}
  \label{fig:xsect_ewkcteq}
\end{figure}

\begin{figure}[htbp]
  \begin{center}
        \includegraphics[width=12cm]{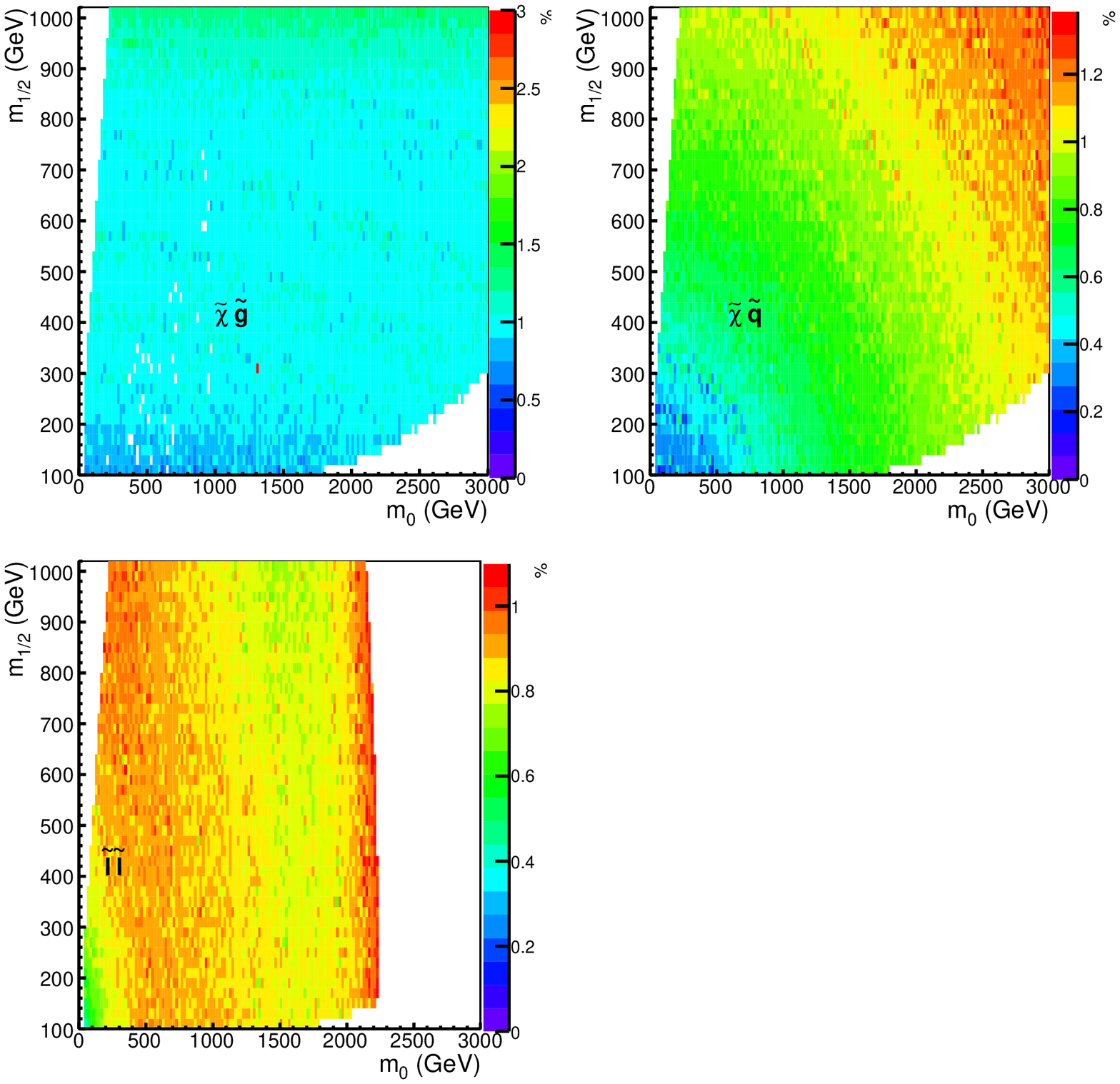}
  \end{center}
  \caption{Uncertainties in \% due to \mstw~PDF error prescription for SUSY electroweak
    productions in the $m_{0} - m_{1/2}$ plane. The CMSSM framework is
    used, with $\tan\beta = 10, A_{0} = 0 $~GeV and $ \mu > 0$. 
}
  \label{fig:xsect_ewkmstw}
\end{figure}

SUSY electroweak particles are usually produced via cascade decays but
they can also be directly produced. In this section we study the
production cross sections along with the theoretical uncertainties
using a CMSSM~\cite{cmssm} scenario with $\tan\beta = 10, A_{0} = 0 $~GeV and $ \mu
> 0$.  Figure~\ref{fig:xsect_ewknom} shows the NLO cross section for
various SUSY electroweak productions in the $m_{0} - m_{1/2}$ mass
plane.  The pure electroweak states have much lower cross sections in
comparison with SUSY colour production. All possible combinations of
charginos/neutralinos ($\tilde{\chi}$) and sleptons ($\tilde{l}$) are
considered for the rate given by $\tilde{\chi} \tilde{\chi}$ and
$\tilde{l}\tilde{l}$ sub-processes, respectively. The theoretical
uncertainty due to the scale variation is found to be less than 10\% for
pure electroweak states as shown in Figure~\ref{fig:xsect_ewkscale}.
The squark and electroweak gaugino associated sub-processes can have
errors as large as 35\%. The systematics due to the \cteq~and
\mstw~PDFs for $\tilde{\chi} \tilde{\chi}$ are found to be less than
5\%.  The uncertainties on the rest of the sub-processes, as shown in
Figure~\ref{fig:xsect_ewkcteq} and Figure~\ref{fig:xsect_ewkmstw}, are
negligible.

\section{Summary and future prospects}
\label{sec:summary}
We presented here a proposal~\footnote{This documentation should be cited along with
  the original papers, i.e.\ Refs.~\cite{Beenakker:1996ch, Kulesza:2008jb, Kulesza:2009kq, Beenakker:2009ha, Beenakker:2011fu}
 for inclusive squark/gluino production, Refs.~\cite{Beenakker:1997ut, Beenakker:2010nq, Beenakker:2011fu} 
for stop or sbottom direct production, and Ref.~\cite{Beenakker:1999xh} for the production of electroweak gauginos.}, based on state-of-the-art higher-order
calculations, for the definition of
benchmark production cross sections and associated theoretical
uncertainties of various SUSY processes of interest at the LHC.  The
common set of prescriptions described in this document aims to
establish a reference framework for the theoretical input used by the
\atlas\ and \cms\ collaborations in the interpretation of their
measurements.

We presented a collection of explicit results for collisions at a
centre-of-mass energy of 7 TeV.  The theoretical systematic uncertainties are
larger for higher masses, and they are typically dominated by the PDF
uncertainties. These have a significant impact when assessing the
experimental constraints or the sensitivity to a given SUSY model.
Detailed numbers and tables for a broad class of SUSY models and
parameters are collected at the SUSY cross section working group 
twiki page~\cite{combined7TeV}, where the results corresponding to higher collision energies will also appear
soon.

The LHC has recently raised the centre-of-mass energy and the experiments are already analysing data from these collisions. 
Hence, the current set of prescriptions will evolve and is expected to adapt to 8 TeV collisions. In particular, the mass ranges for which the NLO+NLL calculations are available 
will be enlarged including usage of the complete {\sc PDF4LHC} prescription.

\section*{Acknowledgements}
We would like to thank W.\ Beenakker, S.\ Brensing, R.\ H\"opker, M.\ Klasen, E.\ Laenen, L.\ Motyka,  I.\ Niessen, M.\ Spira and P.M.\ Zerwas for a fruitful collaboration 
on SUSY cross section calculations. 
This work has been supported by the Helmholtz
Alliance ``Physics at the Terascale'', the DFG SFB/TR9 ``Computational
Particle Physics'', the Foundation for Fundamental Research of Matter (FOM), the Netherlands Organisation for Scientific Research (NWO), 
the  Polish National Science Centre grant, project number DEC-2011/01/B/ST2/03643, and the ERC grant 291377 ``LHCtheory''.  MK thanks the
CERN TH unit for hospitality. SP acknowledges support from the DOE under the grant DOE -FG02-90ER40546.


\end{document}